\documentstyle[aps,prl]{revtex}

\def\be{\begin{equation}}
\def\ee{\end{equation}}
\def\bea{\begin{eqnarray}}
\def\eea{\end{eqnarray}}
\def\sp{{\mathcal C}}
\def\tr{{\rm tr}}

\newcommand{\la}{\langle}
\newcommand{\ra}{\rangle}

\begin{document}

\draft

\title{Statistical distinguishability between unitary operations}

\author{A. Ac\'\i n}

\address{Departament d'Estructura i Constituents de la
Mat\`eria, Universitat de Barcelona, Diagonal 647, E-08028 Barcelona, Spain.\\
e-mail: acin@ecm.ub.es
}

\date{\today}

\maketitle


\begin{abstract}

The problem of distinguishing two unitary transformations, or quantum gates, is analyzed and a function reflecting their statistical distinguishability is found. Given two unitary operations, $U_1$ and $U_2$, it is proved that there always exists a finite number $N$ such that $U_1^{\otimes N}$ and $U_2^{\otimes N}$ are perfectly distinguishable, although they were not in the single-copy case. This result can be extended to any finite set of unitary transformations. Finally, a fidelity for one-qubit gates, which satisfies many useful properties from the point of view of quantum information theory, is presented.

\end{abstract}

\pacs{PACS Nos. 03.67.-a, 03.65.Bz}



\section{Introduction}

Quantum nonorthogonality is one of the basic features of Quantum Mechanics. The deep implications of nonorthogonality can be reflected by the study of the following simple scenario: consider the case in which one has to determine an unknown state chosen from a set of two quantum alternatives that are not orthogonal. It is well known that a complete determination is not possible unless you are provided with an infinite number of copies of the unknown state. Starting from this simple situation, some measures have been defined trying to quantify the degree of orthogonality, or distinguishability, between quantum states, either for pure \cite{Woott} or mixed \cite{BC} states. A geometrical structure for the set of quantum states emerges from these measures: the closer two states, the less distinguishable they are.

Not very much is known about how to extend some of these concepts to the case of unitary operations, although many results were found in \cite{CPR}. In this work, after reviewing some of the existing ideas for quantum states, we look for the measurement maximizing the statistical distinguishability between two unitary transformations. From this result, as it happens for states, one can define a fidelity-like function based on statistical distinguishability which measures the {\em orthogonality} between unitary transformation (or quantum gates). Remarkably, and contrary to what happens in the case of quantum states, it is proved that given two unitary matrices $U_1,U_2\in SU(d)$, it is always possible to find a finite number $N$ such that $U_1^{\otimes N}$ and $U_2^{\otimes N}$ are perfectly distinguishable, although they were not for $N=1$. The case of $SU(2)$ is studied with detail due to its simplicity and importance in quantum information theory. But first, let us review some known results about distinguishability between classical probability distributions, and how they are translated into the quantum domain.

\subsection{Classical probability distributions}

A generic probability distribution of $M$ elements is given by a vector, $\vec p=(p_1,...,p_M)$, with positive components satisfying $\sum_ip_i=1$. The $M-1$ hyperplane generated by these points is called $M$-simplex and corresponds to the space of probability distributions of $M$ elements. There is a privileged metric in it, the Fisher metric, which reads
\be
\label{fisher}
ds^2=\sum_i \frac{dp_i^2}{p_i} .
\ee
It induces a geodesic distance between two probability distributions, $\vec p$ and $\vec q$,
\be
\label{stdist}
d(\vec p,\vec q)=\arccos\left(\sum_i\sqrt{p_iq_i}\right)\equiv\arccos\sqrt F ,
\ee
which can be thought of as a measure of the statistical distinguishability between two probability distributions \cite{Kass}. The square of the term inside the brackets is the overlap or fidelity, $F$, between $\vec p$ and $\vec q$. The Fisher metric is then a measure of distinguishability between two neighboring probability distributions and indeed it is the only metric in the space of probability distributions which is monotone under stochastic matrices \cite{Cenc} (a very natural property any measure of distinguishability should satisfy). Moreover, any generalized relative entropy of the form
\be
H_g(\vec p,\vec q)=\sum_i p_ig\left(\frac{p_i}{q_i}\right) ,
\ee
where $g$ is a convex function on $(0,\infty)$ with $g(1)=0$, and in particular the Kullback information entropy \cite{Kull}, $H_{\log}$, leads to the Fisher metric \cite{LR}.

\subsection{Quantum states}
\label{states}

In \cite{Woott,BC} the classical statistical distinguishability was extended to the quantum domain, for pure and mixed states. Consider the case in which one has to distinguish an unknown given state, chosen from a set of two quantum states, $\rho_1$ and $\rho_2$, belonging to an arbitrary Hilbert space. A measurement will be performed over the system in order to obtain some information about it. The most general measurement in Quantum Mechanics corresponds to a resolution of the identity by means of positive operators, the so called positive operator valued measurement (POVM),
\be
\label{POVM}
\sum_{i=1}^r M_i=1 ,
\ee
with $r$ arbitrary and $M_i\geq 0$. The POVM maps a quantum state, $\rho$, into a probability distribution of $r$ elements with
\be
\label{prob}
p_i=\tr(M_i\rho) .
\ee
The problem of distinguishing the two quantum states is now translated into discriminating between the two probability distributions, $\vec p_1,\vec p_2$ associated to the quantum states through (\ref{prob}). A distance between states is then defined, using (\ref{stdist}) by looking for the measurement apparatus that maximizes the statistical distinguishability between the resulting probability distributions,
\be
\label{qdist}
d(\rho_1,\rho_2)\equiv\max_{M_i}\,\,\arccos\left(\sum_i\sqrt{\tr(M_i\rho_1)\tr(M_i\rho_2)}\right) ,
\ee
which is equivalent to minimize the term inside the brackets, i.e. the fidelity or overlap,
\be
\label{qfid}
\sqrt{F(\rho_1,\rho_2)}\equiv \min_{M_i} \sum_i\sqrt{\tr(M_i\rho_1)\tr(M_i\rho_2)} .
\ee
For the case of one-dimensional projectors, $\rho=|\psi\ra\la\psi|$, Wootters \cite{Woott} proved that (\ref{qfid}) gives
\be
\label{psfid}
F(\psi_1,\psi_2)=\left|\la\psi_1|\psi_2\ra\right|^2 ,
\ee
while for mixed states it was shown in \cite{BC,Fuchs} that the solution of (\ref{qfid}) leads to 
\be
\label{mixfid}
\sqrt{F(\rho_1,\rho_2)}=\tr\left(\sqrt{\sqrt{\rho_1}\rho_2\sqrt{\rho_1}}\right) .
\ee
Both quantities are a measure of the statistical distinguishability between quantum states (it is easy to prove that (\ref{mixfid}) gives (\ref{psfid}) when restricted to pure states). It is remarkable that the fidelity obtained for pure states is equal to the usual overlap, while for the case of mixed states (\ref{mixfid}) is equal to Uhlmann's fidelity \cite{Uhl}, although in principle there was no argument for this coincidence. 

Furthermore the corresponding distance, $d=\arccos\sqrt F$, induces a metric tensor in the space of states based on statistical distinguishability. For pure states one finds the Fubini-Study metric \cite{fub},
\be
\label{Fubini}
ds^2_{ps}=\la d\psi|d\psi\ra-|\la d\psi|\psi\ra|^2 ,
\ee
which is the only metric in the space of Hilbert space rays (pure states without the global phase) invariant under the action of unitary transformations, while for mixed states the statistical distance leads to the Bures metric \cite{bur}. A connection between quantum geometry and statistical distinguishability seems to appear (see also \cite{stgeom}).

\section{The $SU(2)$ case}

Our aim is to extend these ideas to the case of one-qubit gates or $SU(2)$ transformations, looking for a measure of the statistical distinguishability between two unitary matrices, $U_1,U_2\in SU(2)$. After introducing some notation, the strategy that maximizes the statistical distinguishability between two $SU(2)$ transformations is presented. From this result one obtains a measure of their distinguishability, which can be thought of as a fidelity for one-qubit gates.

A generic unitary transformation $U\in SU(2)$ can be parameterized as
\be
\label{su2}
U=\left(\matrix{\cos\theta_1 e^{i\theta_2} && \sin\theta_1 e^{i\theta_3}\cr
-\sin\theta_1 e^{-i\theta_3} && \cos\theta_1 e^{-i\theta_2}}\right) ,
\ee
where $0\leq\theta_1\leq\frac{\pi}{2},0\leq\theta_2\leq 2\pi,0\leq\theta_3\leq 2\pi$ \cite{Corn}. Its spectral decompositions will be denoted by 
\be
\label{uspectr}
U=e^{i\alpha}|u\ra\la u|+e^{-i\alpha}|u^\bot\ra\la u^\bot| ,
\ee
with $0\leq\alpha\leq\pi$ (${\rm det}\,U=1$).

\subsection{Distinguishability of one-qubit gates}

Given two unitary matrices, $U_1,U_2\in SU(2)$, in this section we explore whether it is possible to obtain a fidelity function measuring their statistical distinguishability.
The most general strategy is considered \cite{CPR,nos}: the unitary matrices are applied on one of the qubits of an entangled two-qubit state, $|\psi\ra\in\sp^2\otimes\sp^2$, and we want to find the measurement that maximizes the distinguishability between the states $U_i\otimes 1|\psi\ra\,,i=1,2$. The existing results for pure states \cite{nota} can be used, and from (\ref{psfid}) a fidelity for one-qubit gates is defined as
\be
\label{ufid}
F(U_1,U_2)\equiv\min_{|\psi\ra} |\la\psi|(U_1^\dagger\otimes 1)(U_2\otimes 1)|\psi\ra|^2 .
\ee
The initial state can be written in its Schmidt decomposition,
\be
\label{inst}
|\psi\ra=V\otimes W\left(\cos\omega|00\ra+\sin\omega|11\ra\right) ,
\ee
where $V$ and $W$ are unitary transformations and $0\leq\omega\leq\frac{\pi}{4}$. Since the choice of the second basis is irrelevant, $W=1$ and the expression to be minimized is, with $\rho_A\equiv\tr_B(|\psi\ra\la\psi|)$,
\be
\label{min}
\min_{\rho_A}|\tr(\rho_A U)|^2=\min_{\omega,V}|\cos^2\omega\la 0|U|0\ra+\sin^2\omega\la 1|U|1\ra|^2 ,
\ee
where $U\equiv V^\dagger U_1^\dagger U_2 V$ is again a unitary matrix. Using the parameterization of (\ref{su2}) the quantity to be minimized is equal to
$\cos\theta_1^2(1-\sin^2 2\omega\sin^2\theta_2)$.
The maximal distinguishability, or minimum overlap, is obtained when $|\psi\ra$ is a maximally entangled state, $\omega=\frac{\pi}{4}$, and the fidelity for one-qubit gates reads
\be
\label{sufid}
F(U_1,U_2)=\frac{|\tr(U_1^\dagger U_2)|^2}{4} .
\ee
Note that this expression is independent of $V$ and it is equal to the known trace inner product in the space of square matrices.  

The spectral decomposition (\ref{uspectr}) allows for an alternative derivation of the result which is going to be quite fruitful for its generalization. In fact, writing (\ref{min}) in the basis where $U$ is diagonal, we look for
\be
\label{spmin}
\min_{\rho_A}|{\rm tr}(\rho_AU)|^2=\min_{\rho_{uu}} |\rho_{uu}e^{i\alpha}+(1-\rho_{uu})e^{-i\alpha}|^2
=\cos^2\alpha=\frac{|{\rm tr} U|^2}{4} ,
\ee
where $\rho_{uu}\equiv \la u|\rho_A|u\ra$. All the pure states $|\psi\ra$ such that $\rho_{uu}=\frac{1}{2}$ are optimal for distinguishing two unitary operations satisfying $U_1^\dagger U_2=U$. In particular, it is always possible to find an optimal state, depending on $U$, which is not entangled, while the maximally entangled state is optimal independently of the two gates to be distinguished.

The fidelity (\ref{sufid}) has been also proposed in \cite{nos} and the maximally entangled state of two qubits seems to be the state that best captures the information about one-qubit gates in a single run: it is indeed optimal for the problem of estimating an unknown gate \cite{nos} and, as it has been proved here, for discriminating between two possibles $SU(2)$ operations.


\subsection{$N$ copies}

Consider the situation in which one has to distinguish an unknown one-qubit gate chosen from a set of two alternatives, $U_1,U_2\in SU(2)$, but now $N$ copies of the unknown gate are provided (i.e. it is possible to run the gate $N$ times in parallel). This means that the best strategy maximizing the distinguishability between $U_1^{\otimes N}$ and $U_2^{\otimes N}$ should be obtained. It can be proved that, contrary to what happens for quantum states, there always exists a finite number $N$ such that $U_1^{\otimes N}$ and $U_2^{\otimes N}$ are perfectly distinguishable although this was not the case for $N=1$.

Take as above $U=U_1^\dagger U_2$, with spectral decomposition given by (\ref{uspectr}), with $0\leq\alpha\leq\frac{\pi}{2}$ (when $\frac{\pi}{2}\leq\alpha\leq\pi$ the same reasoning can be applied). The eigenvalues of $U^{\otimes N}$ are $\{e^{\pm iN\alpha},e^{\pm i(N-2)\alpha},...,e^{\pm i(N\,{\rm mod}\,2)\alpha}\}$, where $(N\,{\rm mod}\,2)$ is equal to 1 (0) for odd (even) $N$, with eigenvectors given by the corresponding tensor products of $|u\ra$ and $|u^\bot\ra$. The obtention of the state $|\Psi\ra\in\sp^{2^N}\otimes\sp^{2^N}$, of the composite system AB, minimizing $|\la\Psi|U^{\otimes N}\otimes 1|\Psi\ra|$ will provide us with a measure of the distinguishability between the $N$ copies of the two $SU(2)$ operations. Denoting by $u^N_i$ and $|u^N_i\ra$ the eigenvalues and eigenvectors of $U^{\otimes N}$ and by $\varrho\equiv\tr_B(|\Psi\ra\la\Psi|)$, this quantity can be shown to be equal to (see (\ref{spmin}))
\be
\label{Nmin}
|\la\Psi|U^{\otimes N}\otimes 1|\Psi\ra|^2=|\sum_i \lambda_i\,u^N_i|^2 ,
\ee
where $\lambda_i\equiv\la u^N_i|\varrho|u^N_i\ra$ are positive numbers satisfying $\sum_i\lambda_i=1$. This implies that the optimization of the distinguishability is equivalent to minimize the convex sum of the eigenvalues of $U^{\otimes N}$, which are complex numbers distributed over the circle $|z|=1$ (see also \cite{CPR}). It is now easy to prove that this expression gives zero, i.e. perfect distinguishability, when $N\alpha\geq\frac{\pi}{2}$. Indeed take the first integer number, $N_{min}$, satisfying this condition,
\be
\label{ncopies}
N_{min}=\left[\frac{\pi}{2\alpha}\right] .
\ee
In this case the separable state $|\Psi\ra\equiv |\Psi^s\ra\otimes|0\ra$, where
\be
\label{optst}
|\Psi^s\ra=\sqrt q(|u_{+N}\ra+|u_{-N}\ra)+\sqrt{\frac{1}{2}-q}(|u_+\ra+|u_-\ra) ,
\ee
$u_{\pm N}$ and $u_{\pm}$ are the eigenvectors with eigenvalues $e^{\pm iN_{min}\alpha}$ and $e^{\pm i(N_{min}\,{\rm mod}\,2)\alpha}$, and 
\be
\label{optval}
q=\frac{\cos\left((N_{min}\,{\rm mod}\,2)\alpha\right)}{2\left(\cos\left((N_{min}\,{\rm mod}\,2\right)\alpha)-\cos(N_{min}\alpha)\right)} ,
\ee
allows for a perfect discrimination between the $N_{min}$ copies of the two unitary matrices, i.e. the states $|\Psi^s_i\ra\equiv U_i^{\otimes N_{min}}|\Psi^s\ra$ are orthogonal. Of course, a very similar procedure can be applied when $N>N_{min}$. The minimal number of copies of the unknown gate, $N(U_1,U_2)$, needed for perfect distinguishability is then given by (\ref{ncopies}). Note that this is always possible with a finite number of copies, unless $U=1$, i.e. $U_1=U_2$.

\subsection{Geometric interpretation}

The measure of the distinguishability induces, as in the case of quantum states, a distance in the space of one-qubit unitary operations. Given $U_1,U_2\in SU(2)$, the distance based on their statistical distinguishability is
\be
\label{udist}
d(U_1,U_2)=\arccos\left(\frac{|\tr(U_1^\dagger U_2)|}{2}\right) ,
\ee
with $0\leq d\leq\frac{\pi}{2}$. Using this formula, the minimal number of copies for perfect distinguishability is the first integer satisfying
\be
\label{nperf}
N(U_1,U_2)d(U_1,U_2)\geq\frac{\pi}{2} ,
\ee
i.e. the closer the two gates are, the larger the number $N$ is.

From this distance, a Riemannian metric in $SU(2)$ is found, 
\be
\label{sufub}
ds^2_U=\frac{1}{4}\left(2\,\tr(dU\,dU^\dagger)-|\tr(U^\dagger dU)|^2\right) .
\ee
and this expression, which is very similar to (\ref{Fubini}), using the parameterization (\ref{su2}) for $SU(2)$ matrices reads
\be
\label{sumetr}
ds^2_U=d\theta_1^2+\cos\theta_1^2d\theta_2^2+\sin\theta_1^2d\theta_3^2 .
\ee
This metric has a nice geometric interpretation. A generic $SU(2)$ matrix can be parameterized by two complex numbers, $\alpha=\alpha_1+i\alpha_2$ and $\beta=\beta_1+i\beta_2$,
\be
U=\left(\matrix{\alpha && \beta \cr
-\beta^* && \alpha^*}\right) ,
\ee
satisfying $|\alpha|^2+|\beta|^2=\alpha_1^2+\alpha_2^2+\beta_1^2+\beta_2^2=1$. Thus, any unitary operation can be thought of as a point in a three-sphere. It is easy to see that the Euclidean metric on this three-sphere is equal to (\ref{sumetr}).

Finally, from the expression of the metric one can derive the unbiased probability distribution of one-qubit operations. This probability distribution should be used when there is no a priori knowledge about the one-qubit gate you are given, so there is no preferred region in $SU(2)$ and all its elements are equally weighted. The unbiased probability distribution is proportional to the volume element given by the square root of the determinant of the metric tensor, and from (\ref{sumetr}) it is found
\be
\label{suprob}
f(\theta_1,\theta_2,\theta_3)=\frac{1}{4\pi^2}\sin(2\theta_1)d\theta_1d\theta_2d\theta_3 ,
\ee
which is the expression of the Haar measure \cite{Corn}, as it was expected.


\section{Arbitrary dimension}

In this section we explore the extension of these ideas to the case of arbitrary dimension, i.e. we look for a fidelity function reflecting the statistical distinguishability between two unitary transformations $U_1,U_2\in SU(d)$. As above, the most general strategy consists on taking a bipartite pure state, now $|\psi\ra\in\sp^d\otimes\sp^d$, and applying the unknown transformation, chosen from a set of two alternatives, over one of the subsystems. The pure state minimizing the overlap $|\la\psi|(U_1^\dagger U_2)\otimes 1|\psi\ra|$ will provide us with a measure of the statistical distinguishability between the two unitary operations. Taking the spectral decomposition of $U=U_1^\dagger U_2$, $\{|u_i\ra,u_i\}$, and $\rho_A\equiv\tr_B(|\psi\ra\la\psi|)$, the statistical distinguishability between the two $SU(d)$ transformations is
\be
\label{sud}
\min|\sum_i\lambda_iu_i|^2 ,
\ee
where $\lambda_i\equiv\la u_i|\rho_A|u_i\ra$. The eigenvalues of $U$ are complex numbers of modulus equal to one. Defining by $2\delta$ the minimal arc length in the circle $|z|=1$ such that all the $u_i$ are included in it, it is not difficult to see, generalizing the result of $SU(2)$, that (\ref{sud}) is equal to zero when $\delta\geq\frac{\pi}{2}$, i.e. one is able to distinguish the two unitary transformations. When $\delta<\frac{\pi}{2}$, the best strategy consist on taking the two eigenvalues whose phases are maximally separated on the unit circle \cite{CPR}. The found fidelity, based on statistical distinguishability, is
\be
\label{sudfid}
F(U_1,U_2)=\cos^2 d(U_1,U_2) ,
\ee
where $d(U_1,U_2)=\min(\delta,\frac{\pi}{2})$. Again, the maximal distinguishability can be obtained with a not entangled state. Note that for $SU(2)$, since there are only two eigenvalues, this formula gives (\ref{sufid}) and the state $|\psi\ra$ can be chosen equal to a maximally entangled state of two qubits, independently of the two unitary matrices. In the general case, $SU(d)$, the results are not as simple and the optimal state depends on the two unitary operations to be distinguished. For example, in $SU(3)$, the identity operator $U_1=1$ can be perfectly distinguished from all the set of unitary transformations
\be
\label{example}
U_2=\left(\matrix{0 & \sin\gamma_1e^{i\phi_3} & \cos\gamma_1e^{i\phi_4} \cr  \sin\gamma_2e^{-i(\phi_4+\phi_5)} & \cos\gamma_1\cos\gamma_2e^{i\phi_2} & -\sin\gamma_1\cos\gamma_2e^{i(\phi_2-\phi_3+\phi_4)} \cr  -\cos\gamma_2e^{-i\phi_5} & \cos\gamma_1\sin\gamma_2e^{i\phi_5} & -\sin\gamma_1sin\gamma_2e^{-i(\phi_3-\phi_4-\phi_5)}}\right) ,
\ee
where $0\leq\gamma_i\leq\frac{\pi}{2},i=1,2$ and $0\leq\phi_i\leq 2\pi,i=1,...,5$ \cite{su3}, using the not entangled state $|\psi\ra=(1\quad 0\quad 0)^\dagger$.

Finally, let us mention that again it is always possible to find a finite number $N$ such that $U_1^{\otimes N}$ and $U_2^{\otimes N}$ are perfectly distinguishable, although this was not the case for $N=1$. The formula for this number is the same as (\ref{nperf}), and it is consistent with the defined measure of statistical distance.

\section{Concluding remarks}

In this work we have studied the problem of distinguishing unitary operations starting from the simplest scenario: an unknown unitary operations is chosen from a set of two alternatives, $U_1,U_2\in SU(d)$. Previous results for quantum states have been used and a measure of the statistical distinguishability between $U_1$ and $U_2$ has been found. Contrary to what happens for quantum states, there always exists a finite number $N$ such that $N$ copies of the unknown gate are enough for its complete determination, although this was not possible when $N=1$. As we have shown, the closer the two gates are, the larger the number $N$ is. Indeed we can generalize this result to the case in which the unknown gate belongs to a finite set of $k$ unitary transformations. By performing $k-1$ tests as described above, each test allows to discard one of the alternatives, so a perfect discrimination is again possible after a finite number of gate runs. The pair of gates that are more distant should be chosen in each test, in order to minimize the number of runs.

For the particular case of $SU(2)$ the found measure of statistical distinguishability (\ref{sufid}) has a nice geometrical interpretation and has been also proposed as a good measure of the similarity between gates from the point of view of estimation of an unknown unitary operation \cite{nos}. Indeed, it is also interesting to define a new measure between unitary operations reflecting, instead of their statistical distinguishability, the overlap resulting from their application, i.e. it compares their ability on average to make quantum states orthogonal. The expression for this quantity will be
\be
\label{newfid}
\bar F(U_1,U_2)=\int d\psi\,|\la\psi|U_1^\dagger U_2|\psi\ra|^2 ,
\ee
which for the case of $SU(2)$ leads to 
\be
\bar F(U_1,U_2)=\frac{1}{3}+\frac{2}{3}F(U_1,U_2) .
\ee
Note that (\ref{newfid}) cannot be equal to zero, since this would imply that spin flip was a unitary operation. 

In view of all these results we propose expression (\ref{sufid}) as a fidelity for one-qubit gates, since it captures the notion of statistical distinguishability between two $SU(2)$ transformations in several ways and it has a nice geometrical interpretation. We hope this function will be useful in any context where a figure of merits for one-qubit gates is required.

\section{Acknowledgements}
The author thanks Enric Jan\'e, Jos\'e Ignacio Latorre, Debbie Leung, Llu\' \i s Masanes and Guifr\'e Vidal for many useful comments and suggestions. Financial support from Spanish MEC (AP98) and ESF-QIT is also acknowledged.


\begin{references}

\bibitem{Woott} 
W. K. Wootters, Phys. Rev. D {\bf 23} (1981), 357.

\bibitem{BC}
S. L. Braunstein and C. M. Caves, Phys. Rev. Lett {\bf 72} (1994), 3439.

\bibitem{CPR}
A. M. Childs, J. Preskill and J. Renes, J. Mod. Opt. {\bf 47} (2000), 155, quant-ph/9904021.

\bibitem{Kass}
For an extended review of many of these concepts see R. E. Kass, Stat. Sci. {\bf 4} (1989), 188.

\bibitem{Cenc}
N. N. Cencov, {\sl Statistical decision rules and optimal inferences},  Transl. Math. Mon. {\bf 53}, AMS, Providence (1982); L. L. Campbell, Proc. AMS {\bf 98} (1986), 135.

\bibitem{Kull}
S. Kullback, {\sl Information theory and statistics}, Wiley, New York (1959).

\bibitem{LR}
A. Lesniewski and M. B. Ruskai, J. Math. Phys. {\bf 40} (1999), 5702, math-ph/9808016.

\bibitem{Fuchs}
C. A. Fuchs, PhD. Thesis, quant-ph/9601020.

\bibitem{Uhl}
A. Uhlmann, Rep. Math. Phys. {\bf 9} (1976), 273; see also R. Josza, J. Mod. Opt. {\bf 41} (1994), 2315.

\bibitem{fub}
J. Anandan, Found. Phys. {\bf 21} (1991), 1265.

\bibitem{bur}
D. J. C. Bures, Transl. AMS {\bf 135} (1969), 199.

\bibitem{stgeom}
A. Fujiwara and H. Nagaoka, Phys. Lett. A {\bf 201} (1995), 119; D. C. Brody and L. P. Hughston, Phys. Rev. Lett. {\bf 77} (1996), 2851.

\bibitem{Corn}
J. F. Cornwell, {\sl Group Theory in Physics}, 44-91, Academic Press, London, (1984).

\bibitem{nos}
A. Ac\'\i n, E. Jan\'e and G. Vidal, quant-ph/0012015.

\bibitem{nota}
It is easy to prove that no gain in the distinguishability is obtained using mixed states instead of pure states.

\bibitem{su3}
J. B. Bronzan, Phys. Rev. D {\bf 38} (1988), 1994.


\end{references}
\end{document}